\documentclass[aps,pra,twocolumn,showpacs,preprintnumbers,superscriptaddress,amsmath]{revtex4}
\usepackage{amssymb}

\usepackage{graphicx}
\usepackage{dcolumn}
\usepackage{bm}

\begin{document}

\title{Entanglement of the Hermite-Gaussian modes states of photons}
\author{Xi-Feng Ren \footnote[1]{mldsb@mail.ustc.edu.cn}, Guo-Ping Guo\footnote[2]{harryguo@mail.ustc.edu.cn}, Jian
Li, and Guang-Can Guo}
\address{Key Laboratory of Quantum Information, University of Science and\\
Technology of China, CAS, Hefei 230026, People's Republic of
China\bigskip }

\begin{abstract}
We found that the Hermit-Gaussian(HG) modes of the down converted
beams from the spontaneous parametric down conversion are
quasi-conserved and the generated photon pairs are HG modes
entangled for some special cases. This is valuable for either the
investigation of fundament properties of multi-dimensional
entanglement or quantum information applications.

\end{abstract}

\maketitle

Quantum entanglement is a very important property of quantum
mechanics. It is the foundation of quantum teleportation, quantum
computation, quantum cryptography, superdense coding, etc.
Spontaneous parametric down conversion(SPDC) namely the generation
of two lower-frequency photons when a strong pump field interacts
within a nonlinear crystal. The down-converted photons from SPDC
can be entangled in not only polarization, or spin angular
momentum, but also the spatial mode such as orbital angular
momentum(OAM). Such spatial entanglement occurs in an
infinite-dimensional Hilbert space. Most applications of
parametric down-conversion in quantum systems make use of spin
entanglement. In recent years, the interest in multi-dimensional
entangled states, or qudits, is steadily growing for its promise
to realize new types of quantum communication
protocols\cite{Bartlett00,Bechmann00,Bourennane01}, and its
properties in quantum cryptography better than qubits\cite
{Bechmann00,Bourennane01,BechmannA00,Guo02}. These
theoretical\cite {Arnaut00,Molina02,Torres03,Torres04} and
experiments\cite {Mair01,Vaziri02,Vaziri03,Langford04} works about
qudits are mostly based on OAM of the photons. It has been shown
that paraxial Laguerre-Gaussian(LG) laser beams carry a
well-defined orbital angular momentum\cite{Allen92}, and that the
LG modes form a complete Hilbert space. This provides a promise to
explore quantum states belonging to multi-dimensional spaces in one photon%
\cite{Molina02,Mair01}.

In this paper, we decompose the generated two-photon wave function
using another type of spatial mode, Hermite-Gaussian(HG) modes.
The HG modes decomposition comes naturally from optical cavity
modes. We can generate beams in different HG modes by controlling
the optical cavity modes. The detection of these beams have also
been experimentally realized using computer generated holograms
and single mode fibers\cite{Langford04}. Our result shows that
there are interesting relations between HG modes of the signal and
idler photons. In some special cases, the HG modes are
quasi-conserved, and the down-converted photon pairs are entangled
in HG modes. These HG modes form an infinite dimensional basis and
may be useful for the investigation of multi-dimensional
entanglement state. We use these results to explain the multi-mode
Hong-Ou-Mandel interference experiment\cite{Walborn03} as
validation. Additionally, we propose a teleportation protocol
encoded in HG modes as a simple potential application of this HG
modes entangled states.

The normalized HG mode is given by

\begin{eqnarray}
HG_m^n(x,y,z) &=&\sqrt{\frac 2\pi }\frac 1{\omega (z)}\sqrt{\frac
1{2^{m+n}m!n!}}H_m(\frac{\sqrt{2}x}{\omega (z)})  \nonumber \\
&&H_n(\frac{\sqrt{2}y}{\omega (z)})Exp(-\frac{x^2+y^2}{\omega (z)^2}-i\frac{%
k(x^2+y^2)}{2R(z)}  \nonumber \\
&&-ikz+i(m+n+1)\psi (z))
\end{eqnarray}
where $H_l(x)$ are the associated Laguerre polynomials,

\begin{equation}
H_l(x)=\sum_{k=0}^{[\frac l2]}\frac{(-1)^kl!}{(l-2k)!k!}(2x)^{l-2k},
\end{equation}
and the standard definitions for Gaussian beam parameters are used:

$\omega (z)=\omega _0\sqrt{1+(z/z_R)^2}:$ spot size,

$R(z)=z(1+(z_R/z)^2):$ radius of wavefront curvature,

$\psi (z)=\arctan (z/z_R):$ Gouy phase,

$z_R=\frac 12k\omega _0^2:$ Rayleigh range.

$\omega _0$ is the beam width at the beam waist, $m$ and $n$ are mode
indices. $m+1$ is the number of nodes in $x$ direction(horizontal
direction), and $n+1$ in $y$ direction(vertical direction).

At the beam waist $(z=0)$, the HG mode can be written as

\begin{eqnarray}
HG_m^n(x,y) &=&\sqrt{\frac 2\pi }\frac 1{\omega _0}\sqrt{\frac 1{2^{m+n}m!n!}%
}H_m(\frac{\sqrt{2}x}{\omega _0})  \nonumber \\
&&H_n(\frac{\sqrt{2}y}{\omega _0})\times Exp(-\frac{x^2+y^2}{\omega _0^2})
\end{eqnarray}
In the following calculation, we use this equation for the sake of
simplification.

In the monochromatic and paraxial approximations, the state generated by
noncollinear SPDC can be written as\cite{Monken98,Kwiat95}:

\begin{equation}
\left| SPDC\right\rangle =C_1\left| vac\right\rangle +C_2\left| \psi
\right\rangle
\end{equation}
where

\begin{eqnarray}
\left| \psi \right\rangle &=&\sum_{\sigma _s,\sigma _i}C_{\sigma _s,\sigma
_i}\int \int_Ddq_sdq_i\Phi (q_s,q_i)  \nonumber \\
&&\left| q_s,\sigma _s\right\rangle _s\left| q_i,\sigma _i\right\rangle _i
\end{eqnarray}
The coefficients $C_1$ and $C_2$ are such that $\left| C_1\right| \gg \left|
C_2\right| $, and $C_2$ depends on the crystal length, the nonlinearity
coefficient, and the magnitude of the pump field, among other factors. The
kets $\left| q_j,\sigma _j\right\rangle $ represents one-photon states in
plane-wave modes labeled by the transverse component $q_j$ of the wave
vector $k_j$ and by the polarization $\sigma _j$\ of the mode $j=s$ or $i$.
The polarization state of the down-converted photon pair is defined by the
coefficients $C_{\sigma _s,\sigma _i}$. The function $\Phi (q_s,q_i)$ is
given by\cite{Monken98}

\begin{equation}
\Phi (q_s,q_i)=\frac 1\pi \sqrt{\frac{2L}K}\upsilon (q_s+q_i)\sin c(\frac{%
L\left| q_s-q_i\right| ^2}{4K}),
\end{equation}
where $\upsilon (q)$ is the normalized angular spectrum of the pump beam, $L$
is the length of the nonlinear crystal in the propagation $(z)$ direction,
and $K$ is the magnitude of the pump field wave vector. The integration
domain $D$ is, in principle, defined by the conditions $q_s^2\leq k_s^2$ and
$q_i^2\leq k_i^2$. In this paper, we assume that $\Phi (q_s,q_i)$ does not
depend on the polarizations of the down-converted photons\cite
{Walborn04,Kwiat95}. Then the two-photon wave function after we omit the
dependence on the $z$ coordinate is\cite{Walborn04}

\begin{equation}
\psi (\rho _s,\rho _i)=E(\frac{\rho _s+\rho _i}{\sqrt{2}})F(\frac{\rho
_s-\rho _i}{\sqrt{2}})
\end{equation}
where $E(\rho )$ is the normalized electric field amplitude of the pump beam
and $F(\rho )=\frac{\sqrt{KL}}{2\pi z}\sin c(\frac{KL}{8z^2}\rho ^2)$.

Let us suppose that the nonlinear crystal is illuminated by a HG beam with
mode indices $(m_p,n_p)$. In order to study the relations of the HG\ mode of
the down-converted photons, we will expand the two-photon wave function $%
\psi (\rho _s,\rho _i)$ in terms of the HG basis functions $%
HG_{m_s}^{n_s}HG_{m_i}^{n_i}$:

\begin{equation}
\psi (\rho _s,\rho
_i)=\sum_{m_s,n_s}\sum_{m_i,n_i}C_{m_sm_i}^{n_sn_i}HG_{m_s}^{n_s}(\rho
_s)HG_{m_i}^{n_i}(\rho _i).
\end{equation}
$C_{m_sm_i}^{n_sn_i}$ is given by

\begin{eqnarray}
C_{m_sm_i}^{n_sn_i} &=&\int \int d\rho _sd\rho _iHG_{m_p}^{n_p}(\frac{\rho
_s+\rho _i}{\sqrt{2}})F(\frac{\rho _s-\rho _i}{\sqrt{2}})  \nonumber \\
&&HG_{m_s}^{n_s}(\rho _s)^{*}HG_{m_i}^{n_i}(\rho _i)^{*}
\end{eqnarray}

When $L$ is small enough (the thin-crystal approximation), Eq. (9) can be
rewritten as

\begin{equation}
C_{m_sm_i}^{n_sn_i}\propto \int d\rho HG_{m_p}^{n_p}(\sqrt{2}\rho
)HG_{m_s}^{n_s}(\rho )^{*}HG_{m_i}^{n_i}(\rho )^{*}
\end{equation}
For the special case that the pump beam is a Gaussian beam with waist $%
\omega _p$, which means $m_p=n_p=0$, and the signal and idler beams have the
same waist $\omega _o$, we can write Eq.(10) as:

\begin{equation}
C_{m_sm_i}^{n_sn_i}\propto P_{m_s}^{m_i}P_{n_s}^{n_i}
\end{equation}
where

\begin{equation}
P_m^n=\left\{
\begin{array}{c}
\sqrt{\frac 1{2^mm!2^nn!}}\sum_{j=0}^{[\frac m2]}\sum_{k=0}^{[\frac n2]} \\
\frac{(-1)^{k+j}m!n!2^{m+n-2j-2k}}{(m-2j)!(n-2k)!j!k!}(\frac{m+n-2j-2k-1}2)!
\\
(1+a)^{-\frac{m+n-2j-2k+1}2}:\text{ }m+n\text{ }is\text{ }even \\
0:\text{ }m+n\text{ }is\text{ }odd
\end{array}
\right.
\end{equation}
and $a=(\frac{\omega _o}{\omega _p})^2$. It is easy to find that when $a\ll
1 $, $P_m^m\gg P_m^n(m\neq n)$. If the $x$ index of the signal(idler) beam
is $m$, the probability to find the idler(signal) beam being the same $x$
index is $Q_m=(P_m^m)^2/\sum_{n=0}^\infty (P_m^n)^2$. Fig. 1 gives the
relation between $Q_m$and $a$ for $m=0,1,2$. Fig. 1 shows when $a\ll 1$, the
mode of the signal beam is almost as same as the idler beam. In this
condition, $C_{m_sm_i}^{n_sn_i}$ can be written as

\begin{equation}
C_{m_sm_i}^{n_sn_i}\varpropto \delta _{m_s+m_i}\delta
_{n_s+n_i}P_{m_s}^{m_i}P_{n_s}^{n_i}
\end{equation}
In the similar conditions, we can find for the general case that the pump
beam is in mode $(m_p,n_p)$, the mode of the signal beam $(m_s,n_s)$ and
idler beam $(m_i,n_i)$ have interesting relations:

\begin{equation}
\left\{
\begin{array}{c}
m_s-m_i=\pm m_p \\
n_s-n_i=\pm n_p
\end{array}
\right. .
\end{equation}
We call these relations quasi-conservation laws. It has been experimentally
showed that the condition of $a\ll 1$ can be satisfied with the help of lens
in recent work\cite{wang04}.

Thus we can see that HG modes are quasi-conserved in the SPDC process for
thin crystal. But it should be noted that this conservation can not directly
lead to the HG modes entanglement between down-converted photon pair. This
conservation can also be satisfied by classical correlations between the
generated photon pair. It is well known that the down-converted photons have
LG modes entanglement. In the following, we will show that LG modes
entangled photons can also be HG modes entangled under this
quasi-conservation condition.

The two-photon state generated by the SPDC process is\cite{Mair01,Ren04}

\begin{equation}
\left| \psi \right\rangle =\sum_{l=-\infty }^\infty C_{l,-l}\left|
LG_0^l,LG_0^{-l}\right\rangle
\end{equation}
for $p=0$ if the pump beam is in the $LG_0^0$ mode, where $l$ is the mode
index which represents OAM. HG modes can be expressed as sums of LG modes
and vice versa. The relation between HG mode and LG mode is\cite
{Heckenberg92}

\begin{eqnarray}
\left\langle HG_m^n|LG_p^l\right\rangle &=&\left\langle
LG_p^l|HG_m^n\right\rangle  \nonumber \\
&=&\left\{
\begin{array}{c}
i^mb(\frac{N+1}2,\frac{N-1}2,m):2p+\left| l\right| \\
=m+n \\
0:2p+\left| l\right| \neq m+n
\end{array}
\right.
\end{eqnarray}
where $b(n^{\prime },m^{\prime },m)=\sqrt{\frac{(m^{\prime }+n^{\prime
}-m)!m!}{2^{n^{\prime }+m^{\prime }}n^{\prime }!m^{\prime }!}}\frac 1{m!}%
\frac{d^{m^{\prime }}}{dt^{m^{\prime }}}[(1-t)^{n^{\prime }}(1+t)^{m^{\prime
}}]|_{t=0}$. Combine Eq.(15) and (16), and use the quasi-conversation laws,
we will gain:

\begin{equation}
\left| \psi \right\rangle =\sum_{l=-\infty }^\infty \sum_{m=0}^{\left|
l\right| }g_{m,l}\left| HG_m^{\left| l\right| -m},HG_m^{\left| l\right|
-m}\right\rangle .
\end{equation}
We can see the conservation of HG modes is satisfied by a quantum
correlation (entanglement) but not classical correlation of the
down-converted fields. State (17) can be simplified as:

\begin{eqnarray}
\left| \psi \right\rangle &=&\sum_{m,n}C_m^n\left| HG_m^n,HG_m^n\right\rangle
\nonumber \\
&=&C_0^0\left| HG_0^0,HG_0^0\right\rangle +C_0^1\left|
HG_0^1,HG_0^1\right\rangle  \nonumber \\
&&+C_1^0\left| HG_1^0,HG_1^0\right\rangle +C_1^1\left|
HG_1^1,HG_1^1\right\rangle +......
\end{eqnarray}
where $C_m^n$ is given by $C_m^n\varpropto P_m^mP_n^n$ according to Eq.(13).
The state (18) is a multi-dimensional entangled state. As an example, if we
just consider the first three terms and let $a=0.25$, the state is $%
0.66\left| HG_0^0,HG_0^0\right\rangle +0.53\left| HG_0^1,HG_0^1\right\rangle
+0.53\left| HG_1^0,HG_1^0\right\rangle $. These two photons with HG modes
entanglement can be valuable for either the investigation of fundament
properties of multi-dimensional entanglement or quantum information
applications.

Using the same method, we can find that for the general case(the condition $%
a\ll 1$ is not required), we can just get the entanglement of parity of HG
modes. This means:

\begin{equation}
\left\{
\begin{array}{c}
Parity(m_s+m_i)=Parity(m_p) \\
Parity(n_s+n_i)=Parity(n_p)
\end{array}
\right. .
\end{equation}
This conclusion is also found by Walborn and his co-workers at the same time%
\cite{Walborn04x}.

The entanglement of HG modes can be verified using the quantum state
tomography method\cite{Langford04}. This experiment is currently in progress
in our laboratory.

As validation, we show below that the present theory can be used to explain
the Hong-Ou-Mandel interference experiment\cite{Walborn03}. In this
experiment, the authors state that the transverse field amplitude of the
pump beam has been transferred to the two-photon wave function\cite
{Walborn03,Monken98}. Actually, the experiment result can be directly
explained when the down-converted photons are written in HG modes entangled
states.

If the pump beam is in $HG_0^1$ mode, the generated two-photon state is (we
discard the $x$ index for the sake of simplification\cite{Walborn03}):

\begin{equation}
\left| \psi _0^1\right\rangle =\frac 1{\sqrt{2}}(\left| 1,0\right\rangle
+\left| 0,1\right\rangle )
\end{equation}
where $\left| 1,0\right\rangle $\ means the parity of $y$ index of the
signal beam is even and idler beam odd, and $\left| 1,0\right\rangle $ means
the parity of $y$ index of the signal beam is odd and idler beam even.

The coincidence detection amplitude for an experimental set-up similar to
the work\cite{Walborn03} is given by

\begin{equation}
\Psi _c=\Psi _{tt}(r_1,r_2)+\Psi _{rr}(r_1,r_2)
\end{equation}
where $\Psi _{tt}(r_1,r_2)\propto \frac 1{\sqrt{2}}(\left| 1;0\right\rangle
+\left| 0;1\right\rangle )\Pi (\sigma _1,\sigma _2)$and $\Psi
_{rr}(r_1,r_2)\propto \frac 1{\sqrt{2}}(\left| 1,0\right\rangle +\left|
0,1\right\rangle )\Pi (\sigma _2,\sigma _1)$. $\Pi (\sigma _1,\sigma _2)$ is
the four-dimensional polarization vector of the photon pairs. If $t=r\approx
1/2$ and $\Pi (\sigma _1,\sigma _2)=-\Pi (\sigma _2,\sigma _1)$, $\Psi _c=0$%
, which means there is no coincidence detections; but if $\Pi (\sigma
_1,\sigma _2)=\Pi (\sigma _2,\sigma _1)$, coincidence detections is not
zero. For the case that the pump is in $HG_1^0$ mode, we will get the
contrary result. These are as same as the results of Walborn\cite{Walborn03}.

In the same way, we can find if the two-photon state is $\frac 1{\sqrt{2}%
}(\left| 1,0\right\rangle -\left| 0,1\right\rangle ),\frac 1{\sqrt{2}%
}(\left| 0,0\right\rangle +\left| 1,1\right\rangle )$ or $\frac 1{\sqrt{2}%
}(\left| 0,0\right\rangle -\left| 1,1\right\rangle )$, there will be no
coincidence detections for $\Pi (\sigma _1,\sigma _2)=\Pi (\sigma _1,\sigma
_2)$. This can be used to realize a teleportation protocol encoded in HG
modes. It is different from the early experiment\cite{Walborn03E} in which
the information was encoded in polarization. The idea is that there is a
particle 1 in a certain quantum state, the qubit $\left| \psi \right\rangle $
$_1=\alpha \left| 0_1\right\rangle +\beta \left| 1_1\right\rangle $, where $%
\left| 0\right\rangle $ and $\left| 1\right\rangle $ represent the HG mode
in $y$ direction, and $\left| \alpha \right| ^2+\left| \beta \right| ^2=1$.
Particles 2 and 3 are entangled in the state $\left| \psi \right\rangle $ $%
_{23}=\frac 1{\sqrt{2}}(\left| 1_2,0_3\right\rangle +\left|
0_2,1_3\right\rangle )$. Although initially particles 1 and 2 are not
entangled, their joint HG modes state can always be expressed as a
superposition of four maximally entangled Bell states, since these states
form a complete orthogonal basis. The total state of 3 particles can be
written as

\begin{eqnarray}
\left| \psi \right\rangle _{123} &=&\left| \psi \right\rangle _1\otimes
\left| \psi \right\rangle _{23}  \nonumber \\
&=&\frac 12((\left| 0_1,0_2\right\rangle +\left| 1_1,1_2\right\rangle
)(\alpha \left| 1_3\right\rangle +\beta \left| 0_3\right\rangle )  \nonumber
\\
&&+(\left| 0_1,0_2\right\rangle -\left| 1_1,1_2\right\rangle )(\alpha \left|
1_3\right\rangle -\beta \left| 0_3\right\rangle )  \nonumber \\
&&+(\left| 1_1,0_2\right\rangle +\left| 0_1,1_2\right\rangle )(\beta \left|
1_3\right\rangle +\alpha \left| 0_3\right\rangle )  \nonumber \\
&&+(\left| 1_1,0_2\right\rangle -\left| 0_1,1_2\right\rangle )(\beta \left|
1_3\right\rangle -\alpha \left| 0_3\right\rangle ))
\end{eqnarray}
Now performs a HOM Interference measurement on particles 1 and 2, as we show
previous, if the joint coincidence detections is not zero, particles 1 and 2
is in the state $\frac 1{\sqrt{2}}(\left| 1_1,0_2\right\rangle +\left|
0_1,1_2\right\rangle )$, and then particle 3 is in the state $\alpha \left|
0\right\rangle +\beta \left| 1\right\rangle $. The teleportation protocol is
completed. The experimental setup can be similar to that of Bouwmeester\cite
{Bouw97} while Bell-State measurement is realized by HOM Interference.

In conclusion, we have calculated the probability amplitudes of different HG
modes of the down converted photons generated from SPDC for thin crystal.
Our result shows that the HG modes of the signal and idler beams have some
interesting relations. They can be considered as quasi-conservation for some
special cases. In these cases, the generated two-photon state is a
multi-dimensional state of HG modes. The recent Hong-Ou-Mandel interference
experiment\cite{Walborn03} can be directly explained with the proposed HG
modes entanglement. In addition, we also propose a teleportation protocol
encoded in HG modes as a simple potential application. Our result is
valuable for either the investigation of fundament properties of
multi-dimensional entanglement or quantum information applications..

\begin{center}
\textbf{Acknowledgments}
\end{center}

The Authors would like to acknowledge useful discussion with S. P. Walborn.
This work was funded by the National Fundamental Research Program
(2001CB309300), National Nature Science Foundation of China(10304017), the
Innovation Funds from Chinese Academy of Sciences.

\end{document}